# Monitoring and Proactive Management of QoS Levels in Pervasive Applications

Georgios Boulougaris, Kostas Kolomvatsos

Department of Informatics and Telecommunications

University of Thessaly

Papasiopoulou 2-4, 35131, Lamia, Greece

emails: {gboulougar, kostasks}@uth.gr

**Abstract**

The advent of Edge Computing (EC) as a promising paradigm that provides multiple computation and analytics capabilities close to data sources opens new pathways for novel applications. Nonetheless, the limited computational capabilities of EC nodes and the expectation of ensuring high levels of QoS during tasks execution impose strict requirements for innovative management approaches. Motivated by the need of maintaining a minimum level of QoS during EC nodes functioning, we elaborate a distributed and intelligent decision-making approach for tasks scheduling. Our aim is to enhance the behavior of EC nodes making them capable of securing high QoS levels. We propose that nodes continuously monitor QoS levels and systematically evaluate the probability of violating them to proactively decide some tasks to be offloaded to peer nodes or Cloud. We present, describe and evaluate the proposed scheme through multiple experimental scenarios revealing its performance and the benefits of the envisioned monitoring mechanism when serving processing requests in very dynamic environments like the EC.

## 1. Introduction

The rapid development of the Internet of Things (IoT) contributes to the formation of an infrastructure which produces, treats and consumes huge volumes of data. Numerous ubiquitous devices become the hosts and 'administrators' of the collected data while interacting with end users and their environment to make their lives simpler. The ultimate goal is to have services and data processing mechanisms fully adaptive and responsive to users' demands, especially when we target to offer Pervasive Computing (PC) applications. The dynamic nature of the IoT, which is the result of the autonomy and the heterogeneity of these devices, necessitates novel data processing and administration approaches. Effective interaction models are required to fully unleash their potential. In centralized legacy systems, processing tasks and services occur in the network core, usually present in Cloud. Raw data generated by IoT devices are moved across the network paths in order to be processed in remote data centers. This physical remoteness between data sources and processing mechanisms, however, imposes significant obstacles including an increased latency, limited data processing control, unnecessary resource consumption, safety and privacy vulnerabilities [24], [35]. As a consequence, difficulties arise in maintaining the desired levels of Quality of Service (QoS) as mandated by every application. Apparently, QoS levels may 'dictate' various requirements related to the realization of multiple parameters that govern any specific application. For instance, QoS can be dictated by data, application or networking parameters covering all the aspects of the corresponding application domain.

Edge Computing (EC) comes into the scene as a promising paradigm to develop computation and analytics capabilities in the network edge devices and alleviate the aforementioned problems. The keystone of the EC is that the tremendous quantity of data is suitably processed close to their source, evolving the edge nodes to knowledge producers except for data consumers [24]. One can envision an ecosystem of EC nodes where processing may take place upon multiple distributed datasets. EC nodes can be transformed to intelligent, autonomous entities that may decide, on the fly, their line of actions towards processing the available data and providing responses to various activities having the form of tasks, i.e., requests for the execution of processing upon the distributed datasets.

One can notice the limited resources (compared to the Cloud infrastructure) of EC nodes. To deal with this disadvantage, EC nodes should apply *a selective strategy concerning the tasks that will be executed locally*. For instance, an increased load present in an EC node may force the node to offload some tasks in order to reduce the latency in the provision of the final response. Tasks offloading is a key research topic in EC and IoT if we consider the dynamic environment where nodes act and their heterogeneity [10]. Tasks offloading is first applied to Mobile Cloud Computing (MCC) [6], [22] where tasks are offloaded from mobile nodes to the Cloud back end. The core research challenge is related to the selection of tasks that should be offloaded finding the best possible peer node that will host their execution. The ultimate goal is to keep the execution of tasks at the EC as it can reduce the network traffic [24] driving data analytics towards geo-distributed processing (known as edge analytics) [23], [21], [31]. Evidently, the dynamic environment where EC nodes act imposes various constraints and limitations in the decision making for selecting the tasks that should be offloaded to peer nodes. The variability of the relevant contextual information around the status of *nodes*, *tasks* and the collected *data* may lead to problems concerning the desired levels of QoS.

The main target of our research is the accomplishment, in a distributive manner, of the efficient provision of analytics and the placement of the processing load close to the edge, so as to eliminate the tasks migration to the Cloud, while satisfying of *high QoS levels*. EC nodes decide the appropriate scheduling of the incoming PC tasks based on the available contextual information. Taking into account **(i)** the degree of *data dependency*, **(ii)** the *deadline* of the requested task and **(iii)** the *current load* of each node, EC nodes operate under the ultimate goal of maximizing their performance while minimizing the commitment of resources. The majority of the suggested approaches are characterized as centralized schemes which face the aforementioned problems of Cloud computing. The distributed scheme proposed in this paper is structured upon the autonomous character of EC nodes and intends not only to place the responsibility of scheduling the incoming PC tasks on the network edge but also to meet the requirements for a high QoS. Each node constantly monitors QoS values and proactively has to select some tasks to be offloaded to peers or Cloud. This decision is made when nodes detect that the desired QoS levels are jeopardized (i.e., QoS is estimated to be below a certain threshold). Our decision-making model involves a monitoring mechanism of various performance parameters and a selection strategy of the evicted tasks. We rely on the solution of the widely known *Knapsack problem* [4] and propose a model that aims to maximize the sum of *Degree of Execution* (DoE), i.e., a parameter that depicts if tasks fulfill the resource constraints of each EC node. The

DoE is formulated upon the inference of an *Artificial Neural Network* (ANN) which is fed with a set of parameters representing the surrounding contextual information. We extend one of our previous efforts in the domain (i.e., [18]) by supporting a real time decision making for tasks offloading with the reasoning of the ANN trained upon datasets depicting the appropriate behavior of a node when QoS is jeopardized. The main difference of the current work compared to our previous efforts (e.g., [11], [12], [13], [14], [16], [17]) is the provision of a dynamic scheme for deciding the tasks that will be offloaded paying attention at the 'node level' and not at the 'task level'. Through a 'holistic' process, when QoS is jeopardized, all PC tasks present in the node are evaluated being subject of an offloading action. The novelty of our approach is identified around the following axes:

- The provision of a monitoring scheme for the continuous observation of QoS levels concluded upon a set of performance parameters;
- The incorporation of the communication cost in the decision making when offloading actions are decided, thus, involving the effects on the network before we decide any offloading activity;
- The continuous evaluation of the probability of violating the desired QoS levels resulting a proactive mechanism that prevents any disturbance in the performance of the EC nodes. The envisioned probabilistic approach realizes a quantitative basis through which the relative contribution of QoS violation can be systematically analyzed and support our decision making mechanism;
- The incorporation of a Machine Learning (ML) model, i.e., the aforementioned ANN, to learn the hidden aspects of the execution scenarios for every type of tasks and conclude the DoE which depicts the 'profit' of keeping locally the execution of a task. The proposed ANN is capable of generalizing the decision making being not limited to the provided input.

The following list reports on the contributions of our paper placing it in a set of research domains:

- we provide a distributed decision-making mechanism for PC tasks scheduling which urges the evolution of EC nodes in dynamic environments taking into consideration multiple criteria/parameters;
- we propose a QoS-aware, proactive tasks offloading model upon the continuous monitoring of the performance of EC nodes; we elaborate on comprehensive experimental simulations upon the most significant parameters of the proposed scheme.

The remaining paper is organized as follows. Section 2 reports on the related work while Section 3 discusses the necessary preliminary information and the problem under investigation. Section 4 presents the proposed approach and Section 5 elaborates on the envisioned experiments that reveal the performance of our solution. Finally, Section 6 concludes this paper by presenting our future research directions.

## 2. Related Work

EC promises a significant reduction in the latency for the provision of responses upon task processing performed close to end users. Additionally, it contributes to the efficient bandwidth consumption, while overcomes various resource and energy limitations [24], [35]. As a sequence, there has been an important research interest in production alternative task scheduling schemes.

The decision whether a task could be executed locally at the edge of the network or should be assigned to the remote Cloud, taking into consideration contextual information, seems to be a challenging process. Some of the published research efforts study the task scheduling in the EC taking into consideration a single edge node scenario. Markov chain theory is adopted to analyze the average power consumption and the average delay at the mobile edge device together with an one-dimensional search algorithm proposed by [19]. The final aim is to identify the optimal stochastic task scheduling policy between a mobile device a Mobile Edge Computing (MEC) server. The device energy consumption and the execution delay in a single-user MEC system are also fundamental concepts for the development of a low-complexity sub-optimal minimization algorithm, which uses the flow shop scheduling and convex optimization applications [20]. A sequential task processing approach is presented in [18] to maximize the performance of each node. The proposed scheme includes a pre-processing module for specifying each task's significance with the adoption of the Analytical Hierarchy Process (AHP) and an optimization/selection module for determining the set of tasks which should be locally executed. Those tasks should also satisfy the energy constraints formed upon the solution of the Knapsack Problem [4]. Furthermore, task scheduling, implemented with Greedy Available Fit (GAF) algorithm, resource estimation and virtualization technology, i.e., docker, could be integrated into a gateway-based edge computing service model which aims to fulfill the service requests [26].

Additionally, we can encounter studies that refer to the interactions among numerous edge nodes [3], [5], [29], [33]. These interactions are possible to acquire either in a competitive or in a cooperative nature. Game theoretic approaches for achieving the optimal computation offloading strategy in a distributed manner, while dealing with the competition for the constrained computation and communication resources, are proposed in [3] & [33]. A scheduler may be responsible to look forward to possible cooperation among edge devices in heterogeneous dispersed computing systems to improve execution throughput as depicted by [8]. The scheduler makes use of Directed Acyclic Graphs (DAGs), in which tasks' precedence requirements are determined, and also employs task duplication and task separation as a means of addressing performance bottlenecks. A heuristic, called HealthEdge, is developed in [29] targeting to minimize health tasks processing time. Health tasks can contribute to the diagnosis of possible diseases and to inference of complex behaviors, but at the same time, have increased demands on latency and network bandwidth. HealthEdge chooses the execution destination (the local edge workstation, a remote edge workstation or the Cloud) that offers the least estimation of processing time based both on the task emergency level and the prediction of the human behavior. The collaboration is also desirable in the tiered framework presented in [5]. The proposed task scheduling algorithm, which is delivered upon the Ant Colony Optimization (ACO), exploits the capability of task offloading among IoT devices, cloudlets and cloud datacenters, so that the total difference of the time-dependent profit and the execution cost is maximized for the service provider.

Efficient service or task management satisfying all the specified QoS requirements is a problem which seems to have a significant interest in the EC literature. In particular, an approach which intends to increase the number of the incoming tasks that can be handled by the edge computing network, while complying with the tasks' QoS specifications is developed in [25]. An effort to combine the minimization of the total service provider cost and the maximization of the users' QoS achievement is presented in [27]. Researchers develop two QoS-aware heuristics of placing services and scheduling tasks, one local approach, which treats each device in an independent manner and one global approach, which treats devices as pieces of a wide system, to approximate the optimal solution. Similar targeting is also detected in the study presented in [32]. However, in that case, FOGPLAN, a lightweight framework, is developed to dynamically deploy or release IoT application services on the network infrastructure to meet all the accommodated QoS terms. Subsequently, two periodically executed greedy algorithms, Min-Viol and Min-Cost, whose goals are to minimize the delay violations and the total cost respectively, are designed in order to direct the service provisioning. Finally, a task scheduling algorithm for EC servers, which focuses not only on reducing the average service time error (the difference between the amount of time really assigned to a task and the amount of time that should be assigned) but also on maintaining QoS requirements via resource provisioning, is proposed in [9]. The proposed algorithm, i.e., the K-LZF, intends on providing accurate proportional fairness with constant scheduling overhead even in heterogeneous IoT networks.

## 3. Preliminaries

We consider a set of edge nodes $N = \{n_1, n_2, \ldots, n_{|N|}\}$ which are interconnected and form a graph $G = (N, E)$, where $E$ is the set of edges between them. Each $e_{ik} \in E$ connects two nodes and is subject to a communication cost $cc_{ik} > 0$. Nodes are also connected with a number of IoT devices that report data to them. Apparently, they become the hosts of distributed datasets, i.e., a collection of multivariate vectors setting up the basis for the application of processing mechanisms and knowledge production. The discussed datasets represent the contextual data collected by the environment through the mediation of surrounding end devices and exhibit specific statistical information (e.g., mean, variance or other statistical measures). In Figure 1, we can see an example architecture of the proposed scenario.

**Definition 1**. A dataset is a collection of contextual vectors that depict the information of the environment as observed by recording devices, i.e., $DS = [DV_1, DV_2, \ldots, DV_{|DS|}]$ (|DS| is the number of tuples of the dataset).

**Definition 2**. A data vector is the realization by an observation for a set of dimensions that represent an aspect of a phenomenon under consideration (e.g., temperature, humidity), i.e., $DV = [x_1, x_2, \ldots, x_M]$ (M is the number of dimensions).

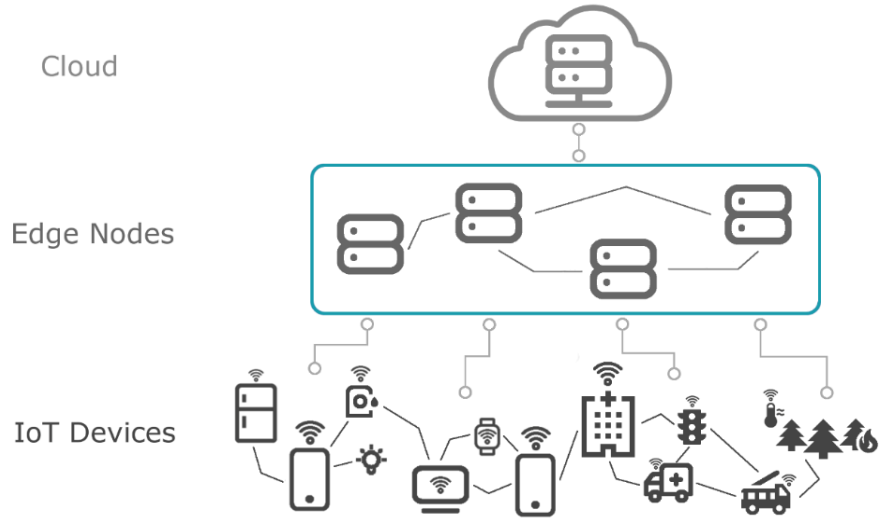

*Figure 1. Example Architecture of the Edge Ecosystem*

Noticeably, the incoming vectors should be pre-processed before they are stored locally. This activity targets to the elimination of outliers, anomalies, etc. that may jeopardize the efficient and objective provision of analytics. The pre-processing phase prepares the data for the upcoming processing and possibly covers their heterogeneity with the adoption of a common model. In this ecosystem, after the aforementioned pre-processing, we consider that nodes exchange the statistical information of the local datasets to inform their peers and facilitate their decision making. Some example activities that mandate the exchange of this statistical information are tasks offloading, data migration to the appropriate nodes and so on and so forth. The frequency of the messaging activities conveying the statistical information of datasets should not be high as the network will be flooded by an increased number of messages. Furthermore, the reporting interval should not be low as peer nodes will not have a `fresh' view on the status of the remote datasets. In any case, in the time interval between two reporting epochs, possible updates may be present in the status of datasets which is a stochastic process, thus, nodes cannot have a clear view on the changes performed in data owned by peer nodes. In the respective literature, one can find efforts dealing with the right time to decide the delivery of updates in order to manage the aforementioned tradeoff (e.g., [15]).

Apart from vectorial data, nodes receive a number of different, independent, single core and interference free tasks (this assumption governs our model) requested by end users or applications. Hence, nodes are responsible to monitor two streams, i.e., the stream which reports the incoming data vectors and the stream that reports the incoming tasks. For each one, a dedicated module is responsible to process the corresponding data or tasks. In this paper, we focus on the latter module and support it with an advanced mechanism for the efficient management of the incoming tasks of pervasive applications. Another source that feeds the stream of tasks deals with tasks offloaded by peer nodes. This is because, nodes may (based on the proposed model) decide to offload some tasks and alleviate their load to meet specific QoS levels. The decision can be made at any point of the execution of the received tasks, however, the interruption of the execution of the current task is not permitted. This means that we do not consider a preemptive approach, i.e., a new task cannot interrupt the execution of the current task. As the current task, we assume the task that is present in the first place of the adopted queue, i.e., all tasks, after their arrival, are placed in a queue which dictates a first-come-first-serve model. The size of the queue and the number of tasks waiting in it define the load of every node. Assume now that $k$ tasks are present in the queue. The current task (i.e., the 1st in the queue) is going to be executed and 'gets' the resources of the corresponding node. The node is monitoring the QoS levels upon multiple parameters (discussed later) and, if needed, decides to select a subset from the L-1 tasks to be offloaded in peer nodes. The rationale is that every node tries to secure the desired QoS levels, thus, it may be useful to offload some tasks to alleviate the current load and minimize the waiting time for each task. Obviously, nodes carefully select a sub-set of tasks based on a tradeoff dealing with the waiting time in the queue and the effect to the local load compared to the time required to offload a task and wait for getting the final response from a peer node. When offloading tasks, the remaining ones will enjoy more quickly the local resources, reducing their total response time.

Nodes should repeatedly follow a set of very simple high level steps and take specific decisions as Figure 2 shows.

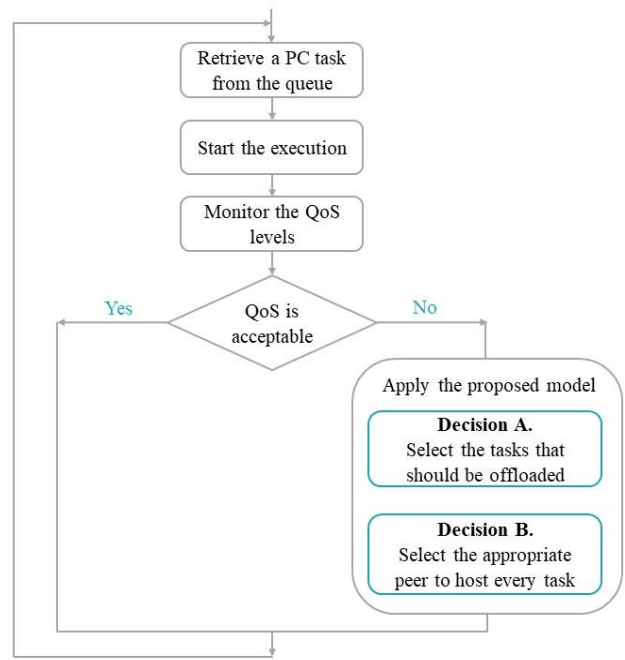

*Figure 2. The flowchart of our monitoring scheme*

Decisions A and B are taken conditioned to if QoS is not at acceptable levels. Then, we have to select the 'appropriate' tasks (based on the model is discussed in the following sections) to be offloaded and to decide the peer node that will host every task. As nodes are connected in a graph, every vertex (node) has a number of neighbors. When tasks are to be offloaded, the 'decision' node selects one from its neighbors to perform the discussed action. The decision is taken upon the statistical information of local datasets and the communication cost $cc$. Past efforts dealing with the selection of the appropriate peer nodes deal with the similarity between tasks' conditions and the available datasets as well as the load that the offloaded tasks will cause to the hosting peers [11], [12], [13], [14], [16], [17]. The novelty of the current approach is the incorporation of the study of the communication cost in the offloading action and the selection of the offloaded tasks based on a study that deals with the probability of QoS violation in order to 'fire' the offloading action. Figure 3 presents the architecture of the proposed model.

At this point, it is important to say that we consider QoS as a measure of the overall performance of an EC node at task level and its ability to provide the best service to the end users and the pervasive applications that it serves. In order to quantitatively measure QoS, we study two aspects of the node's performance, i.e., the *response time* (*RT*) and the *throughput* (*TP*). The response time measures the time required by the specific node to deliver the final response for every task, while the throughput refers to the average number of the served tasks per time unit and exhibits the speed and the ability of the node to quickly deliver the outcomes for multiple tasks. It is necessary our model to demonstrate the lowest response time and the highest throughput, so the highest QoS to be enjoyed not only from the perspective of the end users and the pervasive applications but also from the perspective of each EC node.

**Definition 3**. QoS is the measure of a node's performance that quantitatively describes its ability to provide the best possible service, i.e., $QoS = f(RT, TP)$.

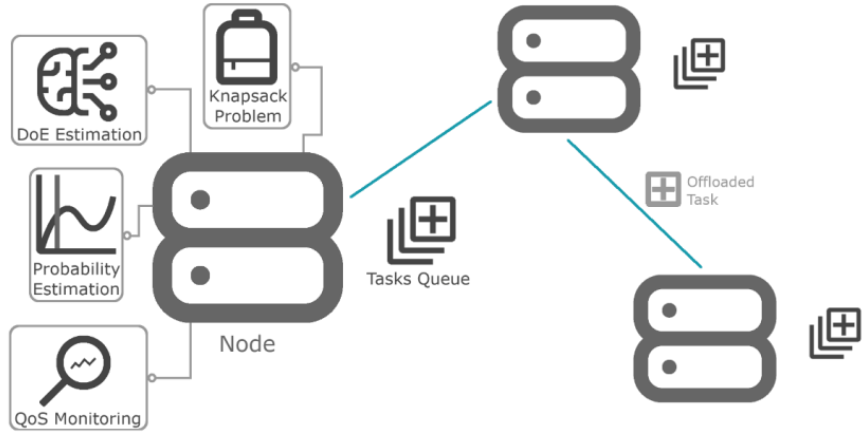

*Figure 3. The Architecture of the Proposed Model*

Nodes are capable of executing a variable set of tasks $T = \{t_1, t_2, ...\}$ depending on the dynamic networked environment and the constraints of their resources. Every task is described by a tuple of characteristics $Ch = \{dd, l, dl\}$ where *dd* is the data dependency, *l* is the load that the task will cause to the host node and *dl* is the deadline of the task. In particular, *dd* depicts the requested data upon which the task should be executed in order to retrieve the corresponding analytics. Usually, *dd* is represented by a set of conditions. For instance, if we consider that a task has the form of a query, *dd* depicts the conditions present in the 'where' clause. For reasons of simplicity, we consider that the task's characteristics take values in the unity interval *I*.

When an offloading action should be present, nodes check their queues and apply the proposed mechanism for selecting the offloaded tasks. Nodes proceed in a tasks' evaluation step based on their characteristics. They feed an ANN with the values of the characteristics and in turn it infers the *Degree of Execution – DoE* $\in [0,1]$. DoE is a measure that depicts the necessity of a local execution. The higher the DoE is, the more imperative is for the corresponding task to be locally executed. The opposite denotes that the local execution is not so critical. As for DoE's value estimation, the combination of a high *dd* value, a low *l* value and a low *dl* value corresponds to a high degree of necessity for task's local execution, while a low *dd* value, a high *l* value and a high *dl* value results in a low DoE value, i.e., the task could be offloaded.

**Definition 4**. DoE is a measure that quantitatively describes to a node how necessary the local execution of a task is $DoE = y(dd, l, dl)$.

Once the DoE has been estimated for each one of the enqueued PC tasks, they, thereafter, are classified in a descending order. The EC node, then, decides which of the tasks will be executed locally and which of them should be offloaded, making use of a suitable mechanism (i.e., the solution of a *0-1 Knapsack problem*) which ensures that the resource requirements will be fulfilled, while simultaneously the sum of the DoE is maximized.

## 4. The Proposed Approach

In this section the proposed scheme and its theoretical foundation is presented in detail. We provide the description of the proposed monitoring mechanism as well as the method for the definition of the DoE.

### 4.1. QoS Modeling and Monitoring

During the execution of PC tasks present in the corresponding queue of each node $n_i$, the node directs the tasks scheduling while monitoring the $QoS \in [0,1]$ estimated values. We consider that QoS should be high, i.e., a value close to unity is desirable to exhibit that the performance of the node is at high levels. $n_i$, during the monitoring process, estimates the probability of having the QoS less than a pre-defined threshold $Th$, i.e., $P(QoS \leq Th)$. This depicts the minimum acceptable performance that every node should exhibit in order to efficiently support the desired tasks. In our model, we consider that the parameters which contribute to the estimation of QoS are: **(i)** the response time ($RT \in [0,1]$) and, **(ii)** the throughput ($TP \in [0,1]$). The list can be easily expanded and incorporate more performance parameters. RT measures, as described above, the amount of time each node takes to respond to a request for a task processing. RT is calculated as the average required time during specific epochs. Each epoch is depicted by a set of discrete steps/tasks executed by the node, i.e., we consider batches of tasks. Let us assume the epoch E where |T| tasks ($\{t_1, t_2, …,\}$) should be executed. $n_i$ records the arrival time for each task and calculates the final response time when every $t_i$ is finalized by producing the final outcome. Assume that the respond time for each task is represented by $rt_i$, then, the following equation holds true:

$$RT = \frac{\sum rt_i}{|T|} \quad (1)$$

In addition, *TP* represents the average number of the served tasks per time unit. The following equation holds true:

$$TP = \frac{|T|}{E_D} \quad (2)$$

where $E_D$ is the total duration of the epoch E.

During its functioning, $n_i$ based on the monitoring of *RT* and *TP* tries to estimate the probability of having QoS over or below the Th, i.e., it tries to calculate the following probabilities:

$$P_{RT} = P(RT \geq Th_{RT}) = 1 - P(RT \leq Th_{RT}) \quad (3)$$
$$P_{TP} = P(TP \leq Th_{TP}) \quad (4)$$

Actually, we adopt two probabilities; one for each QoS parameter, i.e., *Th<sub>RT</sub>* & *Th<sub>TP</sub>*. When the aforementioned probabilities are in place, we are able to aggregate them and deliver $P(QoS \leq Th)$ before we decide to fire the offloading action.

In fact, both *RT* and *TP* are continuous random variables and we have to estimate their probability density function (pdf) and cumulative distribution function (cdf). More specifically, let *RT* and *TP* have the corresponding pdfs $f_{RT}$ and $f_{TP}$, respectively, then, the cdfs $P_{RT}$ and $P_{TP}$ are given by the following equations:

$$P_{RT} = P(RT \geq Th_{RT}) = 1 - P(RT \leq Th_{RT}) = 1 - \int_0^{Th_{RT}} f_{RT}(t) * dt \quad (5)$$

$$P_{TP} = P(TP \leq Th_{TP}) = \int_0^{Th_{TP}} f_{TP}(t) * dt \quad (6)$$

where $Th_{RT}, Th_{TP} \in [0,1]$.

For estimating the discussed pdfs, we rely on the widely known *Kernel Density Estimation (KDE)* [28]. KDE is utilized to deliver the pdf in an unknown random variable. An example execution could be met when we target to improve the prediction accuracy based on a real dataset [34] (the specific dataset contains the performance evaluation of response time and throughput of 5,825 web services by 339 service users). The basic idea behind this approach is that the probability estimation of a given point $x$ is dependent on how many prior points are in $h > 0$ distance units away. Let $(x_1, x_2, \ldots, x_n)$ be finite sample values from an unknown distribution, the pdf of $x$ (its kernel density estimator) can be considered as

$$pdf(x) = \frac{1}{n*h} \sum_{i=1}^{n} K\left(\frac{x - x_i}{h}\right) \quad (7)$$

where $K$ is a non-negative kernel function, like uniform, triangular, normal and Epanechnikov, and $h$ is a smoothing bandwidth parameter. Essentially, a distribution (kernel) is placed on top of each data point $x_i$ and the final distribution estimation arises as the result of the sum of them.

As we focus on a very dynamic environment and target to the edge infrastructure where nodes do not exhibit increased computational capabilities like the Cloud, we decide to adopt an incremental approach for estimating the aforementioned pdfs (then, the cdfs and the envisioned probabilities). We target to adopt W samples for the historical values for RT & TP and deliver the final pdf based on previous estimations. This way, we alleviate the complexity of the approach and save resources.

The pdf $\hat{f}_{RT}(rt_i, j)$ for $j = 1, \ldots, W$ is incrementally estimated by its previous estimate $\hat{f}_{RT}(rt_i, j - 1)$ and the current value $rt_i^j$, that is, we recursively obtain for $j = 1, \ldots, W$ that:

$$\hat{f}_{RT}(rt_i, j) = \frac{j-1}{j*h} \widehat{rt}_i(rt_i, j-1) + \frac{1}{j*h} K\left(\frac{|rt_i - rt_i^{W-j+1}|}{h}\right) \quad (8)$$

If we apply the Gaussian function on the KDE, we obtain an estimation of the cdf $\hat{F}_{RT}(rt_i, W) = \int_{rt_{min}}^{rt_i} \hat{f}_{RT}(u, W) \, du$ using the W values $\{rt_i^{W-j+1}\}_{j=1}^{W}$:

$$\hat{F}_{RT}(rt_i, W) = \frac{1}{W} \sum_{j=1}^{W} \frac{1}{2}\left(1 + erf\left(\frac{rt_i - rt_i^{W-j+1}}{\sqrt{2}}\right)\right) \quad (9)$$

where $erf(.)$ is the error function (also called Gauss error function). Hence, at time j, we obtain the estimation of $\gamma_i = P(rt_i^j > e) \approx 1 - \hat{F}_{RT}(e, W)$. In the first place of our future research agenda is the incorporation of an estimation model for multiple time steps forward than W in our decision making mechanism.

An estimation of the cdf $\hat{F}_{TP}(tp_i, W)$ is respectively obtained by:

$$\hat{F}_{TP}(tp_i, W) = \frac{1}{W} \sum_{j=1}^{W} \frac{1}{2}\left(1 + erf\left(\frac{tp_i - tp_i^{W-j+1}}{\sqrt{2}}\right)\right) \quad (10)$$

The produced probabilities, i.e., one for each performance parameter, are then combined through the adoption of the *Geometric Mean* [7]. The geometric mean is rigidly defined being based on all the desired measurements. Additionally, it gives comparatively more weight to low values, thus, it does not eliminate them from the final result. Finally, the geometric mean has an advantage over the arithmetic mean being less affected by very low or very high values in skewed data. The following equation holds true:

$$P_{QoS} = \frac{P_{RT}^{W_{RT}} * P_{TP}^{W_{TP}}}{P_{RT}^{W_{RT}} * P_{TP}^{W_{TP}} + (1 - P_{RT})^{W_{RT}} * (1 - P_{TP})^{W_{TP}}} \quad (11)$$

where $w_{RT}$ & $w_{TP}$ are the weights for QoS parameter. Evidently, the discussed weights give us the opportunity to pay more attention on one of the parameters, thus, designing a strategy for accepting or rejecting QoS outcomes. For instance, we may want to pay more attention on the response time for each task. In this case, we adopt a high value for $w_{RT}$ eliminating the effect of the throughput in the final probability of QoS violation.

**4.2. Estimating the Degree of Execution**

In case the QoS probability estimation approaches or falls below a certain threshold, node $n_i$ must decide which of the available tasks (Decision A): **(i)** will be locally executed or **(ii)** will be offloaded to the appropriate peer nodes in order a proactive performance improvement to be achieved.

At this point, an evaluation step of the enqueued tasks takes place in EC node $n_i$. $DoE \in [0,1]$ is estimated for each task making use of an ANN [1]. ANNs are abstract computational models inspired by brain neurons structure and function. The adopted ANN is a sequence of functional transformations involving $K$ combinations of the input parameters i.e., $p_1, p_2, \ldots, p_{|M|}$ ($p_k, k = 1, 2, \ldots, |M|$ refers to each task characteristic value). The linear combination of input values can be represented as:

$$a_j = \sum_{k=1}^{|M|} w_{jk} * p_k + w_{j0} \quad (12)$$

where $j = 1, 2, \ldots, C$. In addition, $w_{jk}$ are synaptic weights and $w_{j0}$ are the biases. Subsequently, the activation values $a_j$ are transformed by utilizing the nonlinear sigmoid activation function $g(.)$ to give $z_j = g(a_j)$. The overall output $DoE = y(p)$ that will be the basis for classifying the enqueued tasks can then be obtained from:

$$y(p) = s\left(\sum_{j=1}^{CC} w_j * g\left(\sum_{k=1}^{|M|} w_{jk} * p_k + w_{j0}\right) + w_0\right) \quad (13)$$

where $s(.)$ is the sigmoid function defined as: $s(a) = \frac{1}{1+e^{-a}}$. In the above equation, $CC$ depicts the combinations of the input values and $M$ is the number of the inputs.

We utilize a three-layered feed forward ANN where data related to the realization of *dd* (data dependency), *l* (load) and *dl* (deadline) feed the (first) *input layer*, penetrate the (second) *hidden layer* and end up in the (third) *output layer* with the DoE estimation form. *dd* can be calculated as presented in [14], *l* is depicted by the model proposed in [13], [16] and *dl* is a real value provided in terms of time units (e.g., discrete steps, seconds or any other unit). A critical part of our model is the training process of our ANN. One of our goals is the effective training of the network so as to accomplish a holistic insight in the tasks' characteristics and evaluation. In the training phase, we make use of a training dataset depicting various combinations of input parameters inferring the DoE estimation. The adopted dataset is defined by domain experts.

### 4.3. Tasks Offloading and Allocation

After the tasks' DoE estimation, our model includes a mechanism that concludes the enqueued tasks' execution destination, whether the tasks will be locally executed or will be offloaded. The decision-making process is directed by the solution of a *0-1 Knapsack problem*. The Knapsack formulation is adopted as the assignment process of all tasks to be executed on an individual node it may be unattainable. The reason is that the total workload caused by the enqueued tasks could be higher than the capabilities/capacity of the node. For those tasks that fail to be included in the knapsack, offloading to a suitable destination is chosen (i.e., peer node or the Cloud).

Let us consider that the EC node $n_i$ observes an undesirable QoS estimation and keeps in its queue a finite number of $k$ tasks at a given time, each one of them causes load $l$ to the host node. The load essentially indicates the amount of computing resources (CPU and memory usage, storage and so on) required by a desired task to be accomplished. In addition, $n_i$ is subject to a load capacity constraint $L$ which affects its processing capability. As the total resources are limited, each EC node can perform only a portion of the enqueued tasks at a given time. $n_i$ is also able to conclude the $DoE$ for each task, as described above. The solution of the following optimization problem illustrates the tasks that are preferable to be locally executed.

$$Maximize \sum_{j=1}^{k} x_j * DoE_j \quad (14)$$

$$\text{Subject to} \sum_{j=1}^{k} x_j * l_j \leq L, x_j \in \{0,1\} \quad (15)$$

In the above equations $x_j$ represents the number (restricted to a binary value, i.e., zero or one) of instances of task $j$ to be included in the knapsack. If a task is finally included in the knapsack ($x_j = 1$), its local execution is recommended. Otherwise ($x_j = 0$), the task should be offloaded to a neighboring node or in the Cloud. The solution could be similar to a set of true/false decisions according to whether the current task should be part of knapsack's content aiming to maximize the sum of DoE values. A task is involved in the knapsack by investigating the node's load capacity that is untapped. In other words, the current decision is to examine the load that is left over, or equivalently the load requirements of the tasks previously placed in the knapsack. The configured knapsack problem is resolved each time a QoS violation is detected so as to the solution responds to dynamic queue content of tasks of the EC node.

## 5. Experimental Evaluation

In our experimental evaluation, the performance of the proposed methodology is attempted to be investigated. An effort is made to examine whether the in-edge proactive approach is capable of *effectively deciding task scheduling in the dynamic networked environment while ensuring the satisfaction of QoS requirements*. To do so, various performance metrics are provided.

### 5.1. Performance Metrics

In order to assess our methodology performance, our attention is focused on the node's decision correctness as for either task local execution or task offloading with the ultimate goal of QoS securing. The definition of two simple functions, which depend on node's work load, supports us to return a verdict of node's action. With the use of the function

$$lc_{ti} = b * l \quad (16)$$

we are able to estimate the local execution cost of task. $b \in [0,1]$ refers to a constant value and $l$ refers to the aforementioned task characteristic. On the other hand, the function

$$oc_{ti} = b * l + 2 * cc_{ik} \quad (17)$$

predicts the possible task's offloading cost. The right part of the function takes into consideration twice the communication cost ($cc_{ik}$) between node $i$ and the appropriate peer node $k$. The factor of execution cost $b * l$ is common in both cases.

It is also necessary to assess the correctness of the decision for task local execution, or vice versa, for task offloading to a neighboring node, when the estimation of QoS is less than the predefined threshold. The widely known performance metrics *accuracy A, precision P, recall R* and *F-measure F* provide valuable assistance in our endeavor. The metrics are defined as follows:

$$A = \frac{TP + TN}{TP + TN + FP + FN} \quad (18)$$

$$P = \frac{TP}{TP + FP} \quad (19)$$

$$R = \frac{TP}{TP + FN} \quad (20)$$

$$F = \frac{2 * P * R}{(P + R)} \quad (21)$$

where $TP$ represents a True Positive classification - task local execution is correctly chosen in case the $lc$ is less than $oc$, $FP$ refers to a False Positive classification - task local execution is incorrectly decided when the $lc$ is greater than $oc$, $TN$ represents a True Negative classification - task offloading is correctly chosen in case $lc$ is greater than $oc$ and finally $FN$ refers to a False Negative classification – task offloading is incorrectly decided when $lc$ is less than $oc$.

The selection of the candidate node to which the possible task offloading takes place (Decision B), is done in two alternative ways. In the first case, the node with the lowest work load is preferred, while in the second case, the node that incurs the lowest task's offloading cost is chosen.

In addition, we compare our proposed model with three baseline approaches, regarding the way a model decides the tasks to be offloaded by a node:

- the Random Tasks Offloading (*Random*) model [2] randomly selects a percentage of the enqueued tasks to be offloaded without considering any contextual information;
- the Last Tasks Offloading (*Last*) model offloads the last tasks introduced in the queue;
- the Greedy Highest Load Tasks Offloading (*Greedy*) model [30] selects the tasks that burden the current node load the most to be offloaded.

In all the above cases, the peer node with the lowest load is the preferable destination for task offloading.

In order to support the comparative performance analysis, we form the $C$ metric which depicts the cost of tasks' execution and offloading in an edge ecosystem during an experiment. The following equation holds true:

$$C = \frac{\sum_{i=1}^{\lambda_1} lc + 2 * \sum_{j=1}^{\lambda_2} cc}{\lambda_1 + \lambda_2} \quad (22)$$

$\lambda_1$ is the total number of the locally executed tasks and $\lambda_2$ is the total number of the offloaded tasks. Recall that $lc$ stands for the local execution cost of a task and $cc$ is the communication cost between two peer nodes when a task offloading takes place. Consequently, with the intention of improving the reliability of the performance analysis, we adopt the $AC$ metric which corresponds to the average cost $C$ that arises after the completion of numerous experiments. The $AC$ metric is defined as follows:

$$AC = \frac{\sum_{k=1}^{E} C}{E} \quad (23)$$

where $E$ is the number of the experiments. We also define the difference $D_{AC}$ to illustrate the relative difference in the performance between our proposed model and the substitute methods. $D_{AC}$ is estimated by the following equation:

$$D_{AC} = \frac{AC_m - AC_{aa}}{AC_{aa}} * 100\% \quad (24)$$

$AC_m$ refers to the average cost of the proposed model and $AC_{aa}$ is the performance of the alternative approaches (i.e., $aa \in \{Random, Last, Greedy\}$). A negative value of $D_{AC}$ indicates that the model achieves better $AC$ results than the respective approach, while the opposite is accurate when $D_{AC}$ is positive.

## 5.2. Experiment Setup

Aiming to carry out the comparative analysis described above, we perform two sets of experiments. In the first case, we evaluate the proactive decision to select the task's execution destination when QoS is in danger, considering as number of the iterations of the model's algorithm $itrs \in \{100, 200\}$, while in the latter case, when we compare metric values and we additionally evaluate the $AC$ results between our model and the alternative approaches, we set $itrs = 100$. Furthermore, in the latter case, we examine how the metrics values and the $AC$ results are affected when two different ceilings are imposed on the percentage of tasks offloaded by EC nodes in the alternative models, i.e., 10% and 5%. In any case, we adopt $N \in \{5, 10, 20, 50, 100\}$ and we consider $E = 100$.

In addition, we consider that $RT$ and $TP$ have the same effect on the $P_{QoS}$ probability calculation, i.e., $w_{RT} = w_{TP} = 0.5$ and we make use of the Epanechnikov kernel function for the pdf estimation. Finally, during the experimental process we consider the QoS threshold $Th$ as 0.3 and we make use of the constant value $b = 0.5$ for the estimation of local execution cost $lc$ or the possible offloading cost $oc$ of a task.

Table 1 shows the values of the parameters adopted during the experiments.

*Table 1. Values of parameters in experiments*

| Parameters | Values |
|---|---|
| $b$ | 0.5 |
| $cc$ | [0, 1] |
| $dd$ | [0, 1] |
| $dl$ | [0, 1] |
| $E$ | 100 |
| $itrs$ | {100, 200} |
| $l$ | [0, 1] |
| $L$ | [5, 10] |
| $N$ | {5, 10, 20, 50, 100} |
| $RT$ | [0, 1] |
| $Th$ | 0.3 |
| $TP$ | [0, 1] |
| $w_{RT}$ | 0.5 |
| $w_{TP}$ | 0.5 |

### 5.3. Performance Evaluation

In this subsection, we present an analytical analysis of the metrics as well as their graphical representation that we previously defined in order to provide the full understanding of the experiments.

i. Proactive Decision of Task's Execution Destination

As Figure 4 shows, our model demonstrates a similar behavior as for the selection of the execution destination of the tasks. The majority of the enqueued tasks are locally executed, while a small percentage of them are offloaded to a neighboring node, regardless of whether the one with the lowest load or the one with which the lowest offloading cost incurs is selected. The number of the locally executed tasks in both cases increases at a faster rate as the number of iterations increases, while the number of the offloaded tasks seems not to be affected in a corresponding manner.

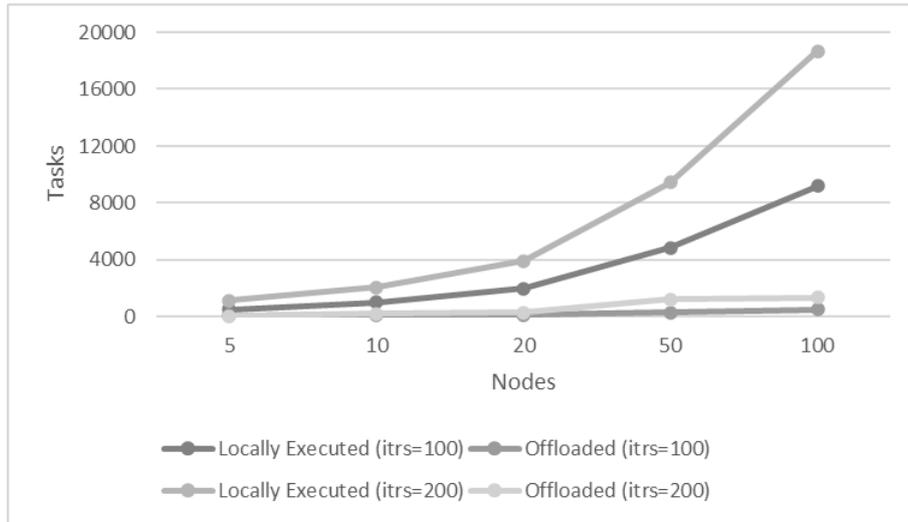

a. Min Load Selection

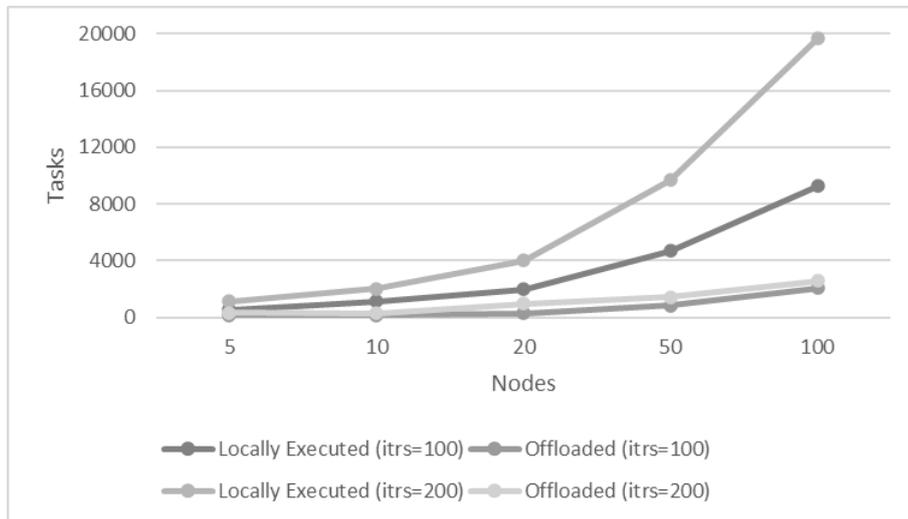

b. Min Task Offloading Cost Selection

*Figure 4. Tasks Destination (with different number of iterations - itrs)*

In Figure 5 and Figure 6, we present the results for the correctness of the decision for task local execution, or vice versa, for task offloading to a neighboring node, when the QoS prediction is less than the predefined threshold. In particular, the metrics *accuracy A, precision P, recall R* and *F-measure F* are shown when the lowest load or the lowest task offloading cost is respectively taken into consideration when selecting the node where the execution takes place, with $itrs \in \{100, 200\}$. Metric values show a remarkable stability in case the necessary task offloading is carried out at the peer node with the lowest load regardless of the number of the participating edge nodes. Our model seems to achieve high levels of efficiency in terms of correctness of decisions. Almost all values are greater than 0.800. The opposite holds true when the pursuit of the lowest task offloading cost is the criterion for selecting the peer node in the event of a possible QoS violation. Except for metric $R$, the values of other metrics tend to decrease as the number of cooperative nodes increases.

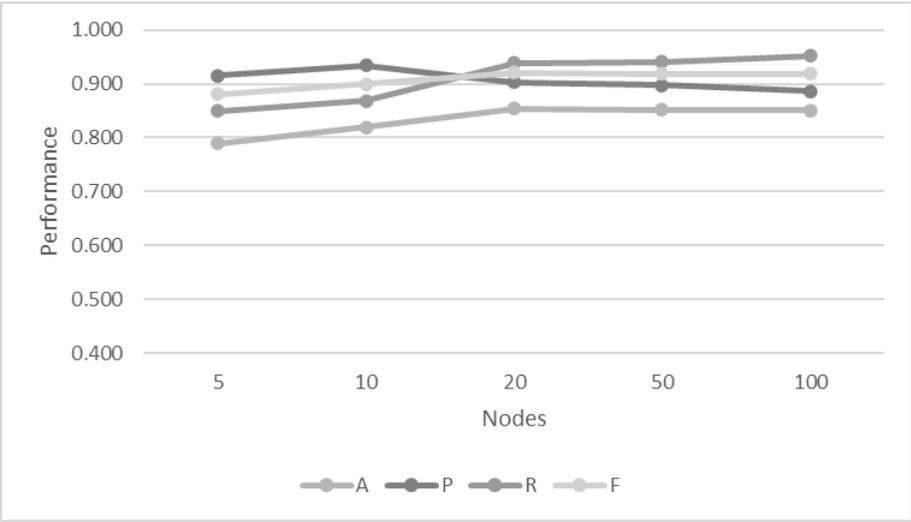

a. itrs = 100

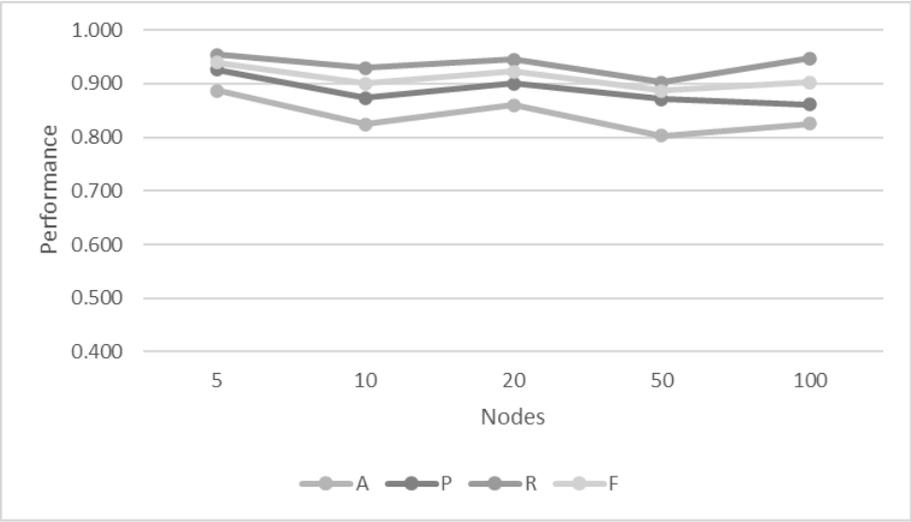

b. itrs = 200

*Figure 5. Metrics - Min Load Selection (with different number of iterations - itrs)*

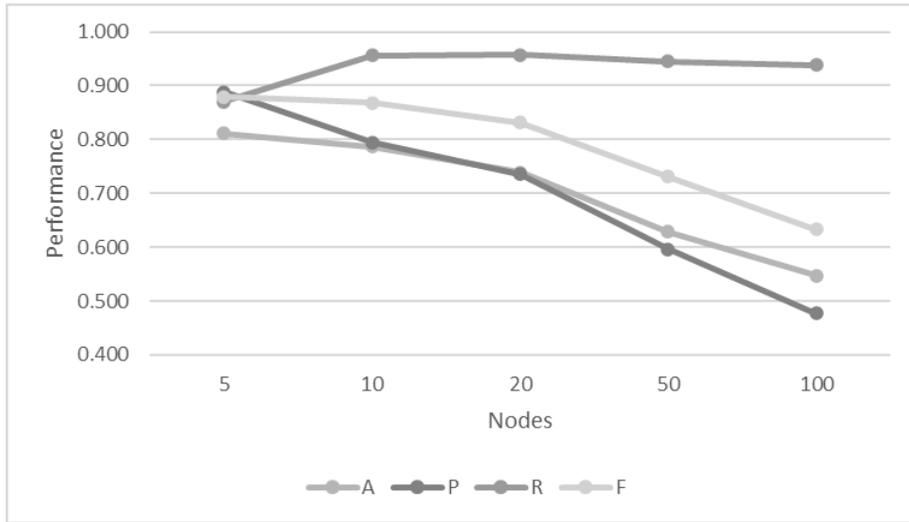

a. itrs = 100

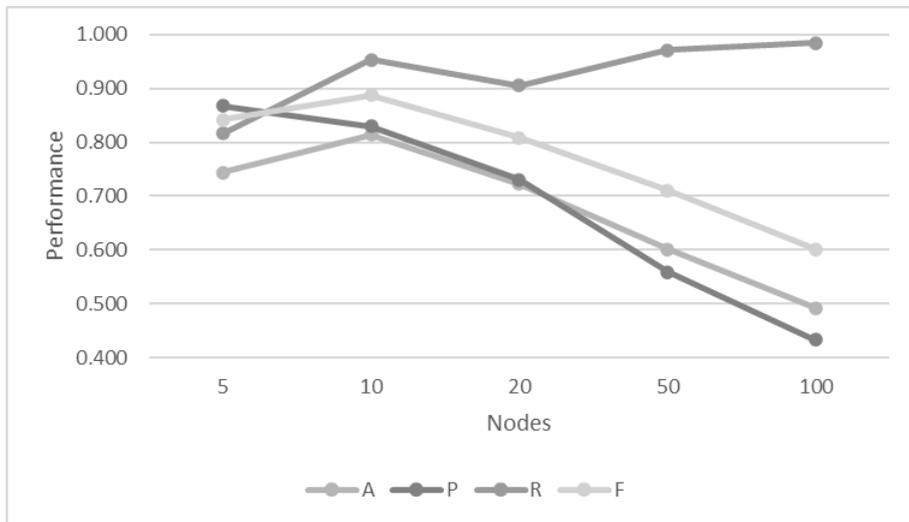

b. itrs = 200

*Figure 6. Metrics - Min Task Offloading Cost Selection (with different number of iterations - itrs)*

ii. Models Comparison

At this point, it is necessary to take into account the percentage of offloaded tasks that results from adopting our model, in order to understand the evolution of the values of its metrics (Table 2). The percentage is getting smaller and smaller as the number of nodes increases.

*Table 2. Model's Percentage of Offloaded Tasks*

| N | Offloaded Tasks (%) |
|---|---|
| 5 | 14.65% |
| 10 | 13.22% |
| 20 | 6.45% |
| 50 | 6.15% |
| 100 | 5.19% |

At the same time, a ceiling is imposed on the percentage of tasks that can be offloaded when considering the substitute methods. We should also recall, before proceeding with the analysis, that the EC node with the lowest load is selected as the task offloading destination when the QoS is in danger of falling below the desired levels. Figure 7 and Figure 8 demonstrate comparisons of the values of the aforementioned metrics between our model and the alternative approaches.

In Figure 7, the maximum percentage of enqueued tasks that can be offloaded from the nodes of the other experimental methods is set as 10%. It is a common finding that metric $A$ exhibits consistently lower values than other metrics among all methods. In the case of our model, the metric $A$ values tend to increase and stabilize at around 0.850 as the number of EC nodes increases, while achieving the highest values of all methods when their number is greater than 20. This means that the proposed model correctly classifies the various instances by about 85% when multiple nodes are involved and collaborating. As for metric $P$, regardless of the selection method, the resulting values appear to slightly decrease as the number of nodes increases. In fact, throughout this experimental process, in most cases, the highest values are recorded in metric $P$. The changes in the values of the metric $R$ are interpreted in a logical way. In case of the proposed model, as the percentage of offloaded tasks decreases, as the number of nodes increases, the corresponding metric $R$ values increase. As for the substitute methods, the $R$ values approach 0.900, due to the imposed ceiling of 10%. With respect to the metric $F$ which is the harmonic mean of $P$ and $R$, we observe that the values of our model tend to increase and reach 0.920, supporting the claim that our model constitutes an effective classification method, while the other approaches end up with values slightly less than 0.900.

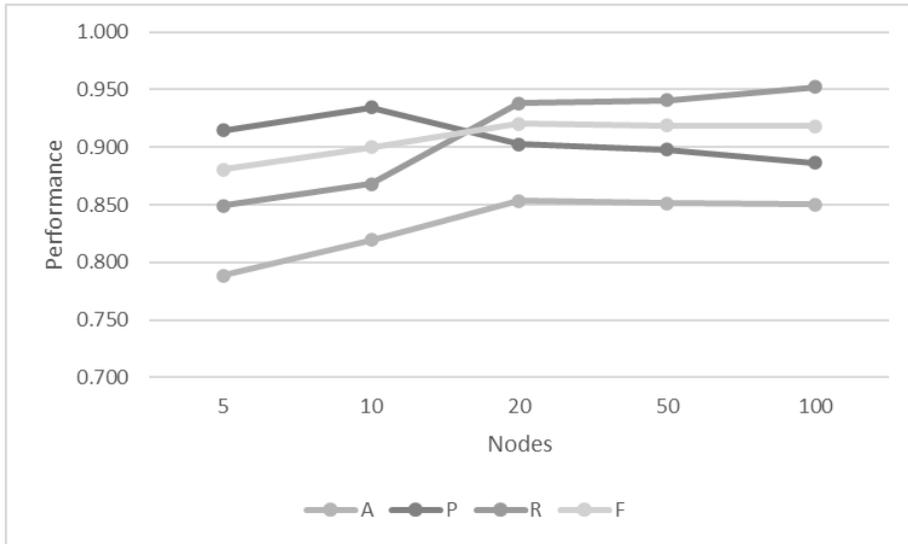

a. Model

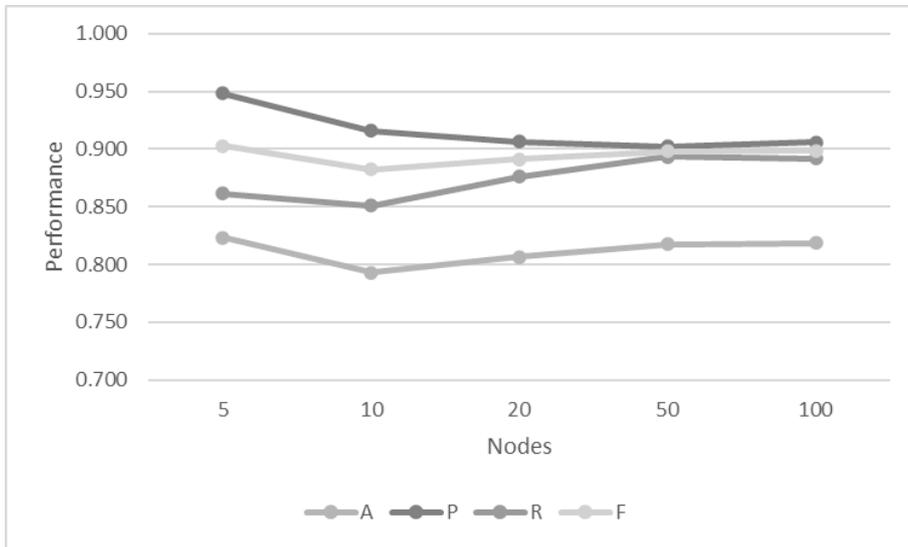

b. Random (10%)

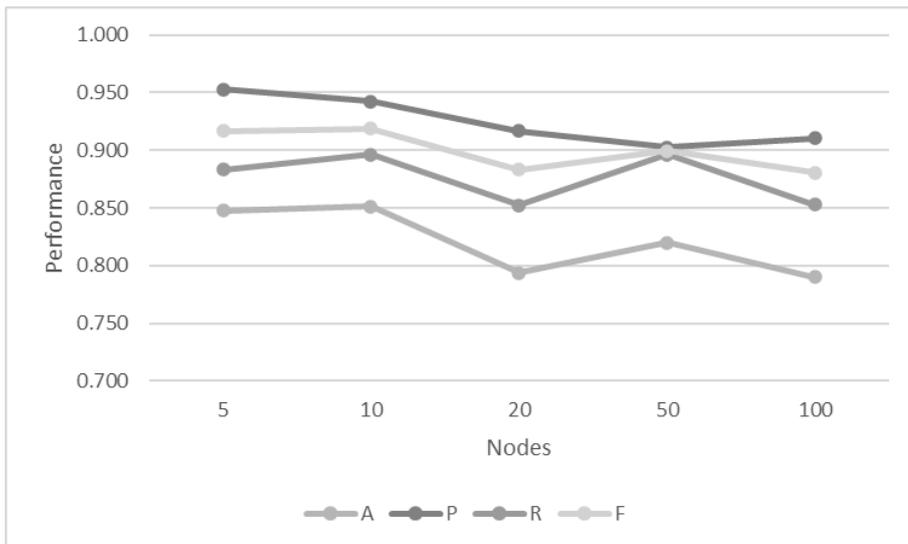

c. Last (10%)

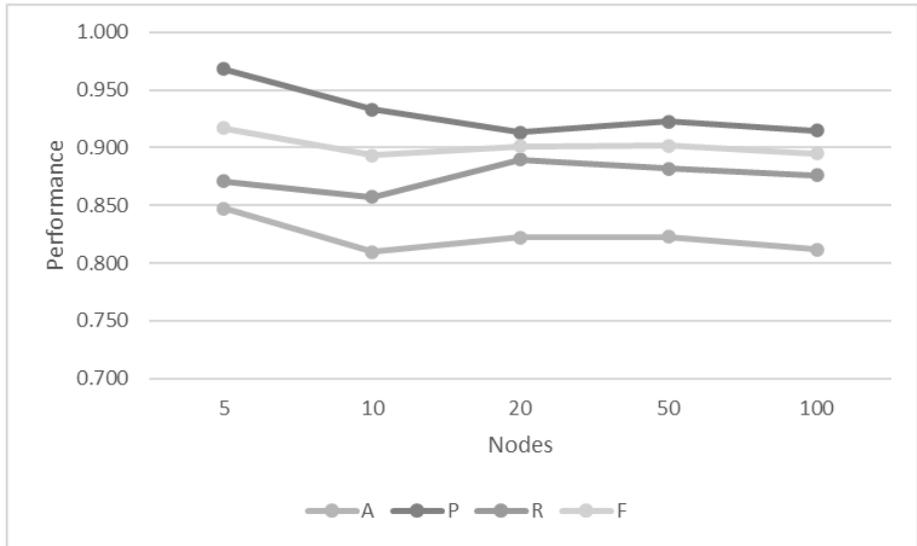

d. Greedy (10%)

*Figure 7. Metrics Comparison*

Figure 8 shows the comparison of metrics when a maximum of 5% of tasks is required to be offloaded from EC nodes, which adopt the substitute methods, in order to meet the QoS requirements. For the sake of convenience, the results of our model are presented again. At first glance, one can observe that the curves of metric *A* and *P* of the alternative approaches are almost identical, slightly moved, while the values of metric *R* form lines that are parallel to the imaginary lines of the values 0.950 and 1.000 due to the very small number of offloaded tasks. *F* outcomes are also affected by the reduction of the percentage of the offloaded tasks, leading to even higher values, range in the interval [0.903, 0.947]. Even in this scenario, however, the proposed model seems to achieve similarly good results compared to the alternative approaches. It is indicative that both *A* and *F* values are slightly higher among all approaches as long as $N > 10$.

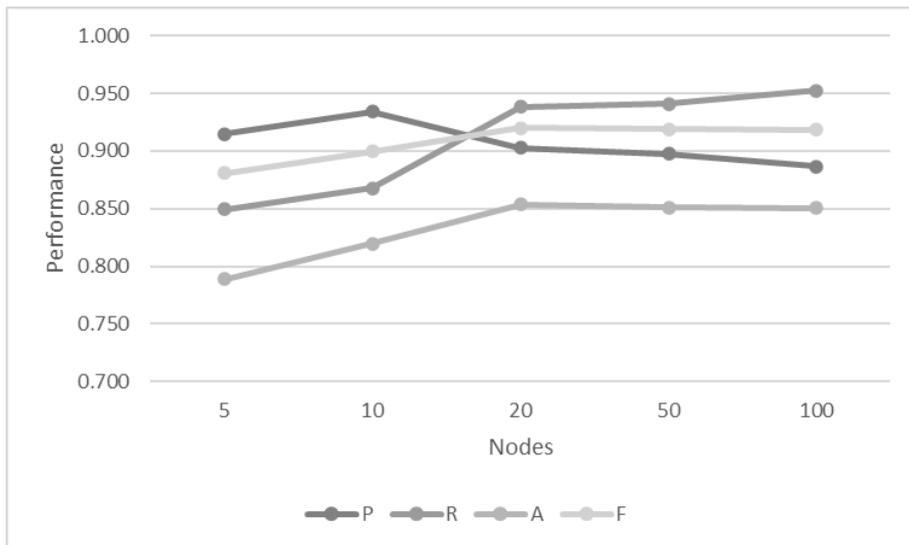

a. Model

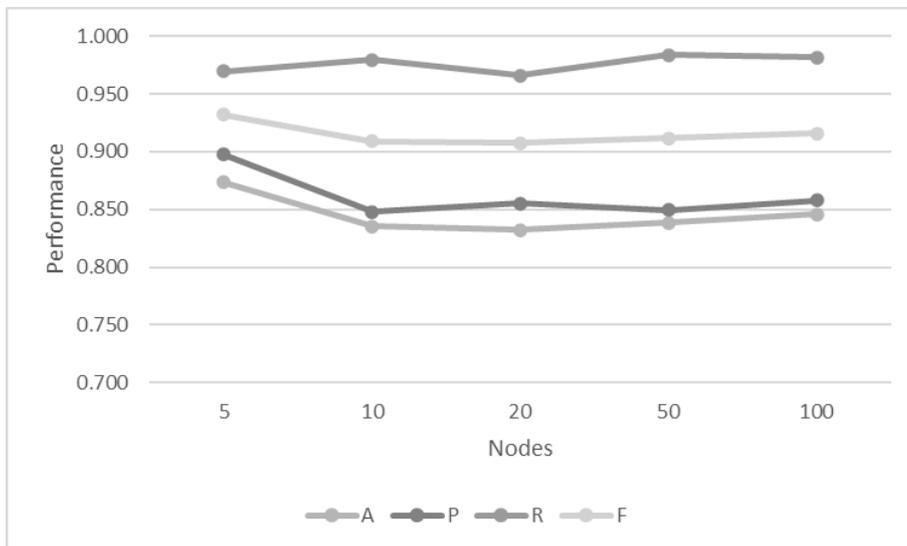

b. Random (5%)

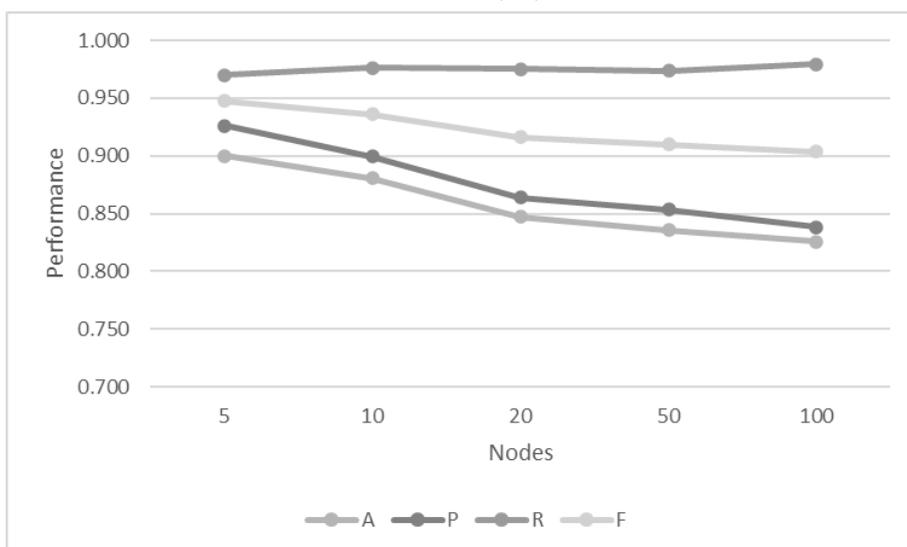

c. Last (5%)

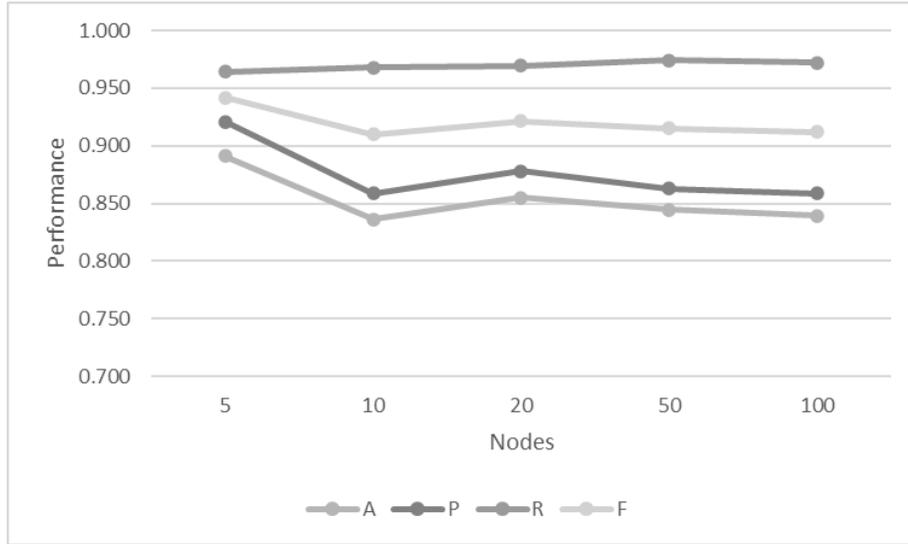

d. Greedy (5%)

*Figure 8. Metrics Comparison*

In Table 3, the aggregated results for the metrics *AC* and *D* are presented. Recall that *AC* refers to the average cost of tasks' execution and offloading and *D* estimates the relative percentage difference in the *AC* between our proposed model and the substitute methods. Two sets of experiments are performed in order to implement the comparative analysis, where, as before, two different ceilings are imposed on the alternative approaches to the percentage of the enqueued tasks that could be offloaded, i.e., 10% and 5%. When a ceiling of 10% is adopted, the proposed model outperforms the *Random*, *Last* and *Greedy* approaches for almost all cases of the particular experimental scenario. Only when the greedy approach is utilized and N = 5, the *AC* (0.435) is less than the corresponding *AC* (0.439) of our model. Respectively, the vast majority of *D* outcomes have a negative sign. The trend is similar in the case of a 5% ceiling. Our model shows better performance than the substitute methods when more than 10 EC nodes are involved. The lower values of *AC* and consequently the positive values of *D*, in case of $N \leq 10$, are largely due to the large difference, of the size of 10%, which is found in the percentage of offloaded tasks. An EC node that adopts the proposed model offloads an average of 14.65% and 13.22% of the enqueued tasks when $N \in \{5, 10\}$ (Table 2).

*Table 3. Comparative Assessment for the Metrics AC and D*

| N | Model | Random (10%) | | Last (10%) | | Greedy (10%) | | Random (5%) | | Last (5%) | | Greedy (5%) | |
|---|---|---|---|---|---|---|---|---|---|---|---|---|---|
| | AC | AC | D (%) | AC | D (%) | AC | D (%) | AC | D (%) | AC | D (%) | AC | D (%) |
| 5 | 0.439 | 0.474 | -7.43% | 0.463 | -5.27% | 0.435 | 0.85% | 0.432 | 1.41% | 0.387 | 13.17% | 0.424 | 3.44% |
| 10 | 0.411 | 0.514 | -20.02% | 0.435 | -5.50% | 0.475 | -13.47% | 0.431 | -4.55% | 0.399 | 3.08% | 0.409 | 0.51% |
| 20 | 0.364 | 0.446 | -18.42% | 0.475 | -23.30% | 0.404 | -9.92% | 0.448 | -18.67% | 0.413 | -11.87% | 0.417 | -12.77% |
| 50 | 0.359 | 0.401 | -10.38% | 0.408 | -11.99% | 0.425 | -15.57% | 0.371 | -3.29% | 0.411 | -12.63% | 0.401 | -10.47% |
| 100 | 0.360 | 0.403 | -10.69% | 0.433 | -16.82% | 0.435 | -17.17% | 0.361 | -0.25% | 0.406 | -11.35% | 0.396 | -9.16% |

Figures 9 and 10 graphically represent the results of the metrics *AC* and *D* in order to contribute to a better understanding of the performance of the different methods.

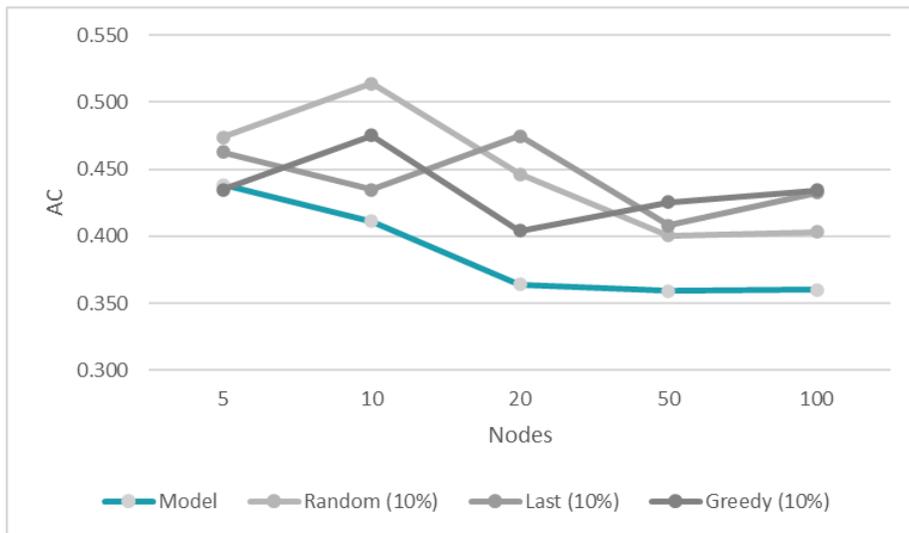

a. Task Offloading 10%

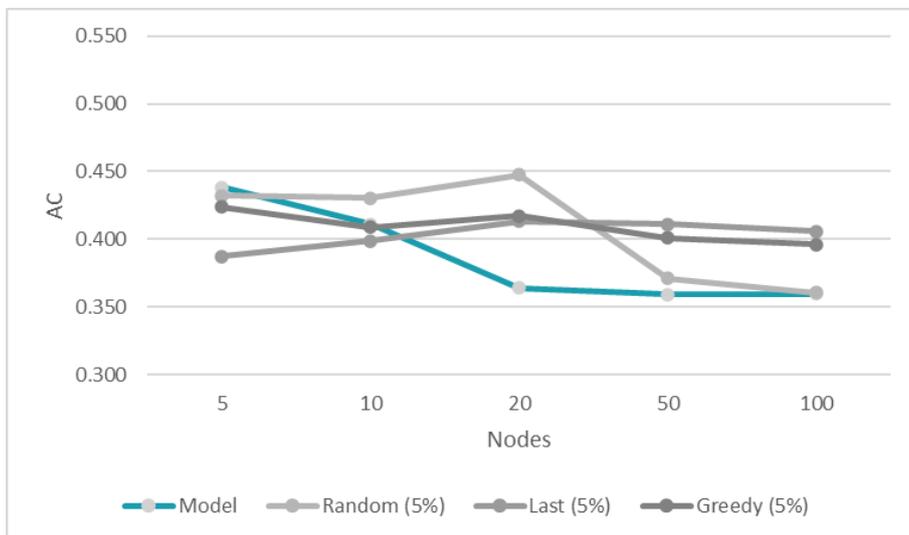

b. Task Offloading 5%

*Figure 9. Average Cost Comparison*

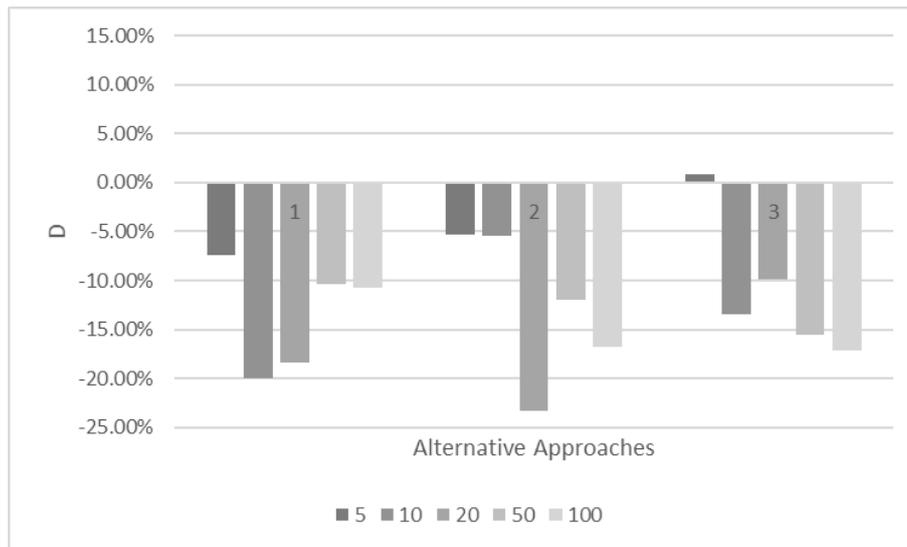

a. Task Offloading 10%

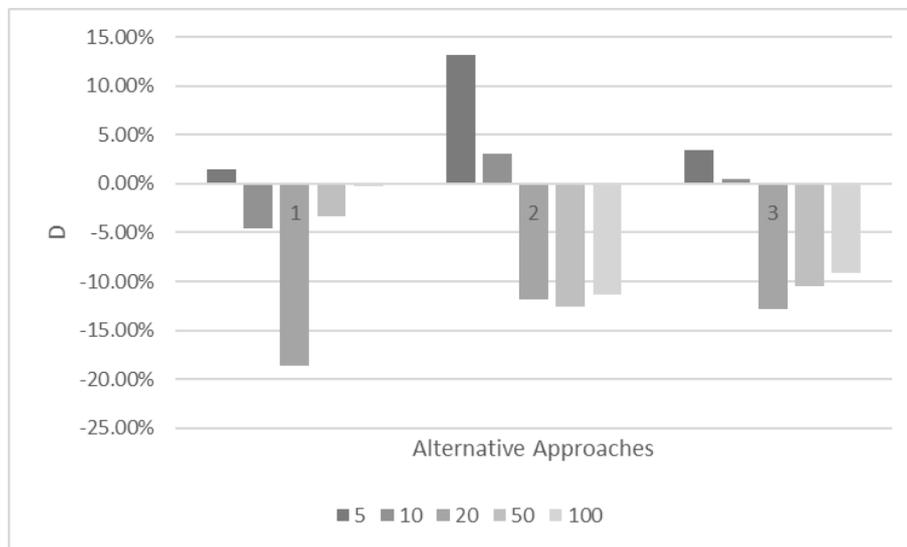

b. Task Offloading 5%

1 - Random   2 - Last   3 - Greedy

*Figure 10. Differences with Alternative Approaches*

Through the aforementioned outcomes and comparisons, we can conclude that the proposed model is capable of exhibiting high quality, in terms of efficiency and performance, decisions in order to proactively fulfill the QoS requirements.

## 6. Conclusions and Future Directions

The evolution of the Edge Computing domain makes nodes capable of executing multiple computation and analytics tasks close to data sources in order to overcome the latency in retrieving results and vulnerabilities related to safety and privacy the centralized systems face. However, maintaining high levels of QoS while addressing the limited computational resources of EC nodes is a challenge. The need arises to formulate an effective selective strategy of the location of tasks execution.

We propose a distributed and intelligent PC tasks offloading model which aims to eliminate the tasks migration to the Cloud, while satisfying of high QoS levels. Each EC node, operating autonomously, systematically observes its performance and it is proactively possible to select some tasks to be offloaded to neighbors or to Cloud, based on their multiple characteristics. The experimental evaluation shows that the proposed model effectively achieves the right decision-making which ensures that resource constraints are met and a high level of QoS is achieved. Future research plans involve the monitoring of the number of hops an incoming task is required to make among the nodes of the EC ecosystem in order to be executed in relation to its available contextual information.